# Origin of Shifts in the Surface Plasmon Resonance Frequencies for Au and Ag Nanoparticles

Sandip Dhara

Surface and Nanoscience Division, Materials Science Group, Indira Gandhi Centre for Atomic Research, Kalpakkam-603 102, India

Email : dhara@igcar.gov.in

## 1. Introduction

Optical properties of noble metal (Au, Ag) nanoclusters have attracted considerable attention in recent years mainly owing to the understanding fundamental issues related to the electronic properties in the small clusters [1] and their applications in nonlinear optics [2], optical switching [3,4], including improved photovoltaic devices [5], cancer therapy [6] and catalysis [7] as well as surface enhanced spectroscopic studies at single molecule level [8-10]. In the present context, the central feature in the optical response is the surface plasmon excitation with coherent oscillation of the conduction electrons which results in a resonance band in the absorption spectra. The other interesting fact about it that for Au, Ag as well as for the other noble metals, the surface plasmon resonance (SPR) occurs in the near-UV-Visible (UV-Vis) region leading to its application with visible light. A large number of experimental and theoretical results were reported for the red or blue shift of the SPR frequency with decreasing noble metal cluster size as an effect of embedding matrix and surrounding porosity [1,11-14] as well as in free noble metal nanoclusters without matrix [15]. Reduction of the electron density (spillout effect) in the small nanoclusters and the screening of the interaction of valence ($s$) electrons by core ($d$)-electrons in noble metals for larger nanoclusters shift the frequency of light absorbed either to the red or blue region, respectively, with decreasing cluster size.

Though strong dependence of the cluster size with the SPR frequency is predicted by time dependent local density approximation (TDLDA) [1], only a few experimental evidences are available for the shift in the SPR frequency for Au clusters with different sizes embedded in alumina [12] and MgO [16,17] matrices without any elucidation to the shift as an effect of size. As a matter of fact, few studies failed to report any shift in the SPR frequency of Au in silica [18-20]. Effect of porosity surrounding Au clusters is taken into consideration in the TDLDA calculation to support the experimental observation [1,13]. A red shift, originating from the spillout effect with increasing polarizability in the system, was predicted by TDLDA calculation for embedded Au clusters in alumina [13].The study in fully embedded Au nanoclusters in crystalline dielectric matrix showed first experimental evidence of red shift of the plasmon frequency with decreasing cluster size for clusters with ~157-427 number of Au atoms [21]. Experimental observations of blue shift of SPR frequency was also reported for Ag [1] and Au [12] clusters embedded in porous alumina matrix and in case of Au clusters (>5 nm) embedded in silica matrix [22]. Most recently quantum effect is also envisaged to understand blue shift of the SPR frequency of individual Ag nanoclusters with reducing size in the range of 2-20 nm [15,23].

## 1.1 Red shift of SPR frequency : Spillout effect

A red shift of the SPR frequency, measured using the UV-Vis spectroscopy with decreasing cluster size was reported in the small sized Au clusters (diameter ~1.7-2.4 nm) embedded in crystalline alumina matrix [21]. The Au nanoclusters were grown in crystalline alumina matrix using 1.8 MeV $Au^{++}$ implantation at various fluence ranges and subsequent annealing at high temperatures. The results of fully embedded Au nanoclusters in crystalline alumina matrix differ completely from other studies where clusters were grown in porous alumina matrices by co-deposition process using pulsed laser ablation technique [12] and sol-gel process [16].

Dipolar Mie resonance of a spherical metal cluster, $\omega_{SPR}= \omega_P/[2\varepsilon_m + \varepsilon_d(\omega_{SPR})]^{1/2}$ where Drude free-electron plasma frequency $\omega_P = (n^2/\varepsilon_0 m_e^*)^{1/2}$ with $n$ as the electron density, $\varepsilon_0$ as the dielectric function of bulk metal, $m_e^*$ as the effective mass of electron and $\varepsilon_d(\omega) = (1 + \chi^d)$ is the core-electron contribution [$\chi^d$ the interband part of dielectric susceptibility (core $d$ electrons)] to the complex dielectric function of the noble metal $\varepsilon(\omega) = 1 + \chi^s(\omega) + \chi^d(\omega)$ [$\chi^s$ the Drude-Sommerfeld part of dielectric susceptibility (valence $s$ electrons)]. The dielectric function of matrix $\varepsilon_m \approx 3.1$ for crystalline alumina in the relevant energy range. The Mie frequency in the large-size limit can be approximately calculated by solving

$$\varepsilon(\omega) + 2\varepsilon_m = 0 \quad \ldots\ldots\ldots\ldots\ldots\ldots\ldots\ldots\ldots\ldots\ldots\ldots\ldots\ldots\ldots\ldots\ldots\ldots\ldots\ldots\ldots (1)$$

in the classical limit, and resulting in

$$\omega_{SPR}(\infty) = \omega_P/[2\varepsilon_m(\omega_\infty) + \text{Re}(\varepsilon_d(\omega_\infty))]^{1/2} \quad \ldots\ldots\ldots\ldots\ldots\ldots\ldots\ldots\ldots\ldots (2)$$

The Mie frequency in the large-size limit for embedded Au nanoclusters in crystalline alumina is approximately calculated to be ~ 2.3 eV (539 nm) using Eqn. (1). Plasmon frequencies around ~ 2.14-2.21 eV (~579-561 nm) are observed in the UV-Vis absorption study of the post-annealed samples for two different annealing conditions 1273 and 1473K (Fig. 1) indicating formation of Au nanoclusters [21]. Cluster sizes were calculated to be in the range of ~1.72-2.4 nm (157-427 number of Au atoms = 32.6-45.4 atomic unit, a.u.~ 0.0529 nm) using acoustic phonon confinement model in low frequency Raman spectroscopic (LFRS) studies [24]. The LFRS study for clusters in the small size range (< 10 nm) were performed for the determination of the size and the shape of clusters. Low-frequency Raman modes in the vibrational spectra of the materials arises due to the confined surface acoustic phonons in metallic or semiconductor nanoclusters. The Raman sensitive spheroidal motions are linked with dilation and strongly depend on the cluster material through $v_t$, the transverse and $v_l$, longitudinal sound velocities. These modes are characterized by two quantum indices $l$ and $n$, where $l$ is the angular momentum quantum number and $n$ is the branch number. $n = 0$ represents the surface modes. The surface quadrupolar mode ($l = 2$, eigen frequency $\eta_2^s$) appears in both the parallel and perpendicularly polarized Raman scattering whereas, the surface spherical mode ($l = 0$, eigen frequency $\xi_0^s$) appears only in the parallel polarized configuration. Considering the matrix effect in the limit of elastic body approximation of small cluster (core-shell model) [25], eigen frequencies for the spheroidal modes at surface ($n = 0; l = 0, 2$) of Au nanocluster in alumina matrix is calculated to be $\eta_2^s = 0.84$ and $\xi_0^s = 0.40$. The surface quadrupolar mode frequencies corresponding to $l = 0$, and 2 are expressed by,

$$\omega_0^s = \xi_0^s v_l / Rc \ ; \ \omega_2^s = \eta_2^s v_t / Rc \ .................(3)$$

where $v_l = 3240$ m/sec, $v_t = 1200$ m/sec in Au and $c$ is the velocity of light in vacuum. Average cluster radii ($<R> \approx 0.86$ -1.2 nm $\approx$ 16.3-22.7 a.u.), calculated using Eqn. (3), corresponding to $l = 0, 2$ are found to be nearly the same conforming to the indirect technique used for the calculation of very small clusters [21]. For both the annealed samples a clear red shift of the SPR frequency was observed with decreasing fluence. For these spherical clusters in the smaller size range, a red shift was observed with decreasing cluster size (Fig. 2a). Though the red shift effect was predicted to be quenched in case of free Au clusters [1], the large matrix-induced charge screening lead to a large electron spillout in case of embedded cluster [13]. A red shift is observed for very small clusters because of the reduction of average electron density with an increasing electron spillout effect with decreasing cluster size. The physics of the blue shift trend observed in noble metal clusters (discussed in the subsequent section) is based on the assumption that because of the localized character of the core-electron wave functions, the screening effects are less effective over a surface layer inside the metallic particle. Close to the surface, the valence electrons are then incompletely embedded inside the ionic-core background. In two-region dielectric model, this hypothesis is taken into account by assuming that the effective polarizable continuous medium responsible for the screening does not extend over the whole cluster volume, where $\chi^d(\omega) = \varepsilon_d(\omega) - 1$ vanishes for radius $> R-r$; where $R$ is the cluster radius and $r$ is a thickness parameter of the order of a fraction of the nearest-neighbor atomic distance (Fig. 2b).

Plasmon frequencies fall in between the values calculated using TDLDA method with $r = 0$ and 1 a.u. in the specific size limit. The red shift trend with decreasing cluster size is clearly depicted for $1>r\geq0$ as the trend is reversed for $r>1$ with increased screening effect of $s$ and $d$ electrons [13]. Normally spillout effect is realized in very small clusters where electron density reduces from its bulk value. Charge density for the clusters upto 170 atoms also showed deviation from its bulk value in the density functional theory (DFT) based LDA calculations applied to the spherical jellium-background model (SJBM) [26]. The SJBM at an infinite flat surface was implemented for high symmetrical potential applicable to significantly large cluster size. However, charge densities and self consistent potentials for 170 atoms vary in their values across the infinite flat surface due to the nature of electronic level filling in the spherical potential. The solution to the problem by introducing lattice structure via pseudo-potentials was limited in the presence of high lattice symmetry for specific elements. The effect of spin using local spin-density approximation was found to be important in this context. However, it was the first effort to address optical response related issues in the DFT-LDA formalism. The model was later on improved up to 440 atoms in the TDLDA calculation (Fig. 3) [13]. In the two-region dielectric model, in addition to the electron density beyond the radius $R$, the magnitude of the red shift was found to depend also on the surface electron density profile. It is analogous to SJBM model [26], involving localized pseudo-potentials. Larger red shift was anticipated with softer surface electron density, as estimated for dielectric function $\varepsilon_d$ in the range 5-10, and was compared the results with those of the standard SJBM. The surface profiles of the mean-field potential and electron-density were found quite identical when $\varepsilon_m = 1$. It suggested similar magnitude of red shift of the plasmon frequency for embedded system as compared to that for free clusters. With number of atoms, $N \sim 427$ (<440) for largest cluster of radius $R \sim 1.2$ nm and 157 for the smallest cluster of $R \sim 0.86$ nm (~ 16.3 a.u.), in the report [21], definitely support the claim that the observed red shift of SPR frequency with decreasing size is due to electron spillout effect in the small Au clusters fully embedded in crystalline alumina matrix. Unlike co-ablated growth of Au in alumina [12], the effect of porosity leading to the blue shift of SPR frequency with decreasing cluster size was proposed to be negligible in our study as the samples were grown in a well crystalline alumina matrix. Carrier escape by interband tunneling, as observed in semiconductor heterojunction nanostructure where electron dynamics of localized Wannier–Stark states [27,28], was reported in the presence of electric field [29,30]. In this context dephasing of electron leading to a dynamic localization of electron in metallic nanoclusters, however was not considered as the coherence lifetime of electron (limited by scattering centers, e.g., point defects and surface states in the nanocluster) was expected to be much smaller than the period of oscillatory motion of electron for all reasonable value of applied electric field.

## *1.2 Blue shift of SPR frequency : Screening effect*

The optical response of free and matrix-embedded Au nanoclusters was reported in the framework of the TDLDA [13]. The characteristics of the SPR frequency with cluster size showed strong influence of the frequency-dependence of the 5*d* core-electron dielectric function in the vicinity of the interband threshold. The size evolution of the Mie-frequency in free Au nanoclusters exhibited a prominent blue shift with decreasing cluster size than that for Ag nanoclusters (Fig. 4). Experimental data corresponding to five different particle-size distributions were shown on composite films consisting of low-concentration (~7%) of Au nanoclusters embedded in an amorphous alumina matrix (black squares) [12]. The experimental data were positioned between the extrapolated theoretical curves corresponding to free and fully-embedded clusters (empty triangles), suggesting that the experimental results might be explained by taking into account the matrix porosity. Neglecting volume change in these nanoclusters and with a mean porosity of about 45% with respect to crystalline $Al_2O_3$ two kinds of model calculations were carried out to understand and quantify the effect of matrix-porosity in these systems. Results obtained for $\varepsilon_m = 2$, assuming a fine-grained homogeneous porous matrix, showed a value corresponding to a decrease of the matrix polarizability by a factor 2 in the visible spectral range ($\varepsilon_m$ ~3.1) (crosses in Fig. 4). In spite of the quantitative agreement for large clusters, the slope of the size evolution was significantly underestimated and the predicted porosity effects might be too small for medium and small clusters. As a matter of fact $\varepsilon_m = 2$, corresponding to a large value of matrix porosity, can be assessed by calculating the dielectric function of the porous alumina matrix using the Bruggeman effective medium theory [31]. The experimentally-determined dielectric function of porous alumina films showed a decrease of ~17% with respect to that for crystalline in the visible spectral range ($\varepsilon_m$ ~ 2.6). At the cluster matrix interface the porosity is important because of the fact that the short-scale heterogeneity of the porous matrix in the vicinity of the interface will play a crucial role as the changes in the induced electron density by the oscillating external field are mostly confined to the surface. Model calculations involving a perfect vacuum "rind" were carried out for $\varepsilon_m = 1$ in the radial range $R < r < R+d_m$. The size evolutions over the entire studied size range were displayed (black triangles in Fig. 4). For a given mean porosity, matrix porosity effects were found stronger in this model. With the slope of the curves in good agreement with the experimental size trend, the deviation from the experimental data was minimized with a vacuum-rind thickness $d_m$ (<< Fermi wavelength, $\lambda_F$). Assessment with recent experimental data obtained with composite films of $Ag_N$ and alumina (black circles in Fig. 4) provides evidence for both the relevance of the porosity effects and suitability of the two-region dielectric model for Au and Ag. The results of TDLDA calculations on $Ag_N$ clusters embedded in alumina, involving the values $d_m = 2$ a.u. and $d_m = 4$ a.u., were plotted (empty circles in Fig. 4). An agreement between theory and experiment, for both Au and Ag, in using the same model parameters, proved the correctness of the model for investigating the size effects in the location of the surface plasmon frequency. In case of matrix embedded noble metal nanoclusters, however the blue shift trend is largely reduced. Agreement with recent experimental results on size-selected Au clusters embedded in an alumina matrix was achieved by taking into account the porosity effects at the metal and matrix interface. The SPR frequencies of Ag nanoclusters, calculated in the TDLDA framework, were in good agreement (Fig. 5) with experimental data on free $Ag_N^+$ clusters [32] and on $Ag_N$ clusters embedded in solid Ar [33]. The size effects were found weak and very sensitive to the matrix, the porosity at the interface, and the cluster charge. It is shown that the core-electron contribution to the metal dielectric function is mainly responsible for the quenching of the size effects in the optical response. In a phenomenological two-region dielectric model for the thickness of the layer of reduced polarizability, the blue shift trend is observed for $r>1$ as the cluster radius decreases (Fig. 2b). Thus, depending on the experiment, a blue shift [34], or a quasi-size-independent evolution [22,35] is exhibited.

The surface plasmon energy of Au nanoclusters formed by $Ar^+$ ion beam mixing of Au/silica was investigated for the size effect [22]. Core-electron contribution to the metal dielectric function was mainly responsible for the blue shift with decreasing cluster size in case of nanocluster diameter >5 nm (Fig. 6a). This is because of the fact that at $\omega_{SPR} > \omega_{max}$ (Fig. 6b) a reduction of $Re[\varepsilon_d(\omega)]$, residing in the denominator of the Eqn (2), with increasing $\omega$ will further enhance the blue shift. The effect was depicted as quenching of size effect in the optical response [1]. The nonzero contribution of $Im[\varepsilon_d(\omega)]$ in Au (Fig. 6b) was reported to broaden the resonance peak by coupling with the interband transition [12]. The broadening of the resonance peak for Au clusters were also reported earlier [11,18,19]. A blue shift of SPR frequency with reducing cluster size was also reported for Au-Ag alloy nanoclusters grown by laser ablation of an alloy target Au-Ag 1:1 atomic composition in the size range of 1.9–2.8 nm [36]. The alloy nanoculsters were embedded with a low concentration in an alumina matrix. Theoretical calculations in the frame work of the TDLDA, including an inner skin of ineffective screening and the porosity of the matrix, were in good agreement with experimental results (Fig. 7). In the standard Mie theory, with a single interface no size effects occur in the dipolar regime ($\lambda>>R$), except for a mere volume scaling factor. Finite-size effects result when the surface skins (thickness $d_m$) of ineffective screening are included in the embedded-cluster model. However, these results are noticeably different from those calculated within the TDLDA-based quantum model (Fig. 7a). This discrepancy originates from the large value of the $\lambda_F \approx 3.3 r_s$ (Wigner radius, $r_s$) relative to $d_m$. The comparison with experiment has been achieved by attributing to the experimental Mie-band maximum the size corresponding to the mean diameter. A noticeable difference was observed between theory and the experiment for large $Ag_n$ clusters. The asymptotic values correspond to a metal sphere embedded in a porous alumina matrix in the experimental findings. Comparing the experimental data of the pure alumina sample with the theoretically calculated value, it was found that the SPR frequency of mixed $(Au_{0.5}Ag_{0.5})_n$ clusters was slightly overestimated in the model. The qualitative analysis, however was quite applicable.

## *1.3 Blue shift of SPR frequency : Quantum effect*

A blue shift of SPR frequency for free Ag nanoclusters is described as quantum plasmonic property for the first time by Scholl *et al.* [15]. They have excited plasmons for individual particles using a scanning tunneling electron microscopic tip where fast electrons are focused on Ag nanoparticles precisely from surface to the central region. It may be noted that the interaction between electromagnetic wave corresponding to fast electrons and plasmons is equivalent to the interaction of light with plasmons. They could differentiate bulk- and surface-plasmon components and a blue shift in the frequency of surface plasmons with decreasing cluster size (Fig. 8). The classical electrodynamics does not predict size dependence of the SPR frequency in the free Ag cluster [11]. The conduction electrons respond to electromagnetic fields as a classical electron gas for nanoparticles with size > 10 nm. In this size range, the contribution of each electron to the plasmons cannot be observed and the SPR frequency is uncertain because of collisions of the electrons with each other and with the atomic lattice of noble metal. However, a clear blue shift was observed for the SPR frequency of the smallest nanoparticles of ~ 2 nm as compared with 10 nm, along with a significant (50%) decrease in the plasmon lifetime. The conduction electrons moving at a Fermi velocity ($v_F$) of ~$10^6$ m/s [21] take approximately 10 seconds to travel across a 10 nm Ag particle. Observed plasmon lifetime of 10 plasmon periods of such particles is close it. However, the blue shift could not be explained using the concept of reflections of the electrons at the surface for the reduction in the plasmon lifetime [11]. The electrons appear at a discrete set of energy levels in nanoparticles smaller than 10 nm, with relatively few conduction electrons (250 electrons for 2 nm nanoclusters) participating in the plasmons. These energy levels are increasingly separated from one another as the particle size is reduced and termed as quantum confinement effect. The individual electron transitions between occupied and unoccupied degenerate electron energy levels increases the uncertainty about the SPR frequency leading to the reduction of plasmon lifetime [37]. In fact, individual transitions cannot in general be resolved owing to the uncertainty produced by the collisions mentioned previously. Considering the effects of individual electron transitions, an analytical quantum mechanical model is described to understand the blue shift due to a change in particle permittivity (Fig. 9). The model also considers catalytic effect, and surface enhanced techniques SERS and TERS effects as manifestation of the quantum plasmon mechanical behavior of tiny noble metal nanoclusters with evanescent field are restricted only to small number of atoms corresponding to the 'hot spot' [38]. Similar observation of blue shift of SPR frequency with decreasing cluster size (Fig. 10), larger than that predicted in the theory, was also reported for isolated spherical Ag nanoclusters dispersed on a $Si_3N_4$ substrate in the diameter range 3.5 – 26 nm [23]. A semi-classical model corrected for an inhomogeneous electron density associated with quantum confinement, and a semi-classical nonlocal hydrodynamic description of the electron density was used to understand the observed blue shift (Fig. 11).

In another report, a blue shift of the quantum SPR frequency with decreasing size (Fig. 12) is reported for 2-10 nm Ag nanoclusters embedded in $SiO_2$ matrix [14]. Reduction in the size with increasing fluence, as explained by the thermal spike model, was achieved by high-energy $Si^{5+}$ ion-irradiations in the embedded Ag nanoclusters substrate. In the noble alloy system optical properties of mixed $(Au_xAg_{1-x})_n$ clusters in the diameter range 1.5–5 nm of various relative compositions embedded in an alumina matrix were also reported [39]. Two simple phenomenological models were developed in estimating the effective dielectric function $\varepsilon_d(\omega)$ of the Au-Ag ionic background, as the optical properties of the mixture depend strongly on the topological non-uniformity in the length scales of the heterogeneities. In one model (Model 1), $\varepsilon_d(\omega)$ of the ionic mixture $Au_xAg_{1-x}$ is assumed to be the composition-weighted average of the interband contributions of bulk Au and Ag metals. The model 1 assumes the mixture as the stacking of small homogeneous Au and Ag grains, whose optical properties are close to those of the corresponding bulk materials. In the second model (Model 2), relations under band structure formalism were derived from a the experimental data obtained from transmission experiments in the energy range of 2.4–4.4 eV [40], and reflection experiments in the energy range of 0.5–6.5 eV [41] on thin homogeneously alloyed Au/Ag films of various compositions. The imaginary components $Im[\varepsilon_d(x,\omega)]$ are plotted for both models (Fig. 13). The prominent differences are (i) the interband threshold is independent of the composition in model 1, while it evolves with $x$ in the second model; (ii) in model 1 the $Im[\varepsilon_d(x,\omega)]$ exhibits a two-step pattern, while a single rising edge is exhibited in model 2, as observed for pure Au and Ag. For a given relative composition, the blue shift of SPR frequency with increasing size was reported to become more important as the Au content increases in the alloy (Fig.14). The size and concentration effects in the optical properties were investigated in the light of quantum effect. These spectra were compared with model calculations, considering quantum mechanical description of $s$ conduction electrons within the TDLDA formalism including both the porosity of the matrix and an inner skin of reduced ionic-core polarizability. Different models of alloy morphology were introduced to describe the effective dielectric function of the ionic background in the noble metal alloy nanocluster. A multi-shell model showed predominant quantum effects. In classical calculations involving a multi-shell cluster of simple metals of different electronic densities, the spectra were structured with as many peaks as the number of interfaces. On the other hand, quantum TDLDA calculations within the jellium model (spillout effect) display only one main resonance peak in the spectra.

**References :**

1. Lermé, J.; Palpant, B.; Prével, B.; Pellarin, M.; Treilleux, M.; Vialle, J. L.; Perez, A.; Broyer, M. Quenching of the size effects in free and matrix-embedded silver clusters. *Phys. Rev. Lett.* **1998**, *80*, 5105-5108.

2. Liao, H. B.; Xiao, R. F.; Fu, J. S.; Yu, P.; Wong, G. K. L.; Sheng, P. Large third-order optical nonlinearity in Au:SiO$_2$ composite films near the percolation threshold. *Appl. Phys. Lett.* **1997**, *70*, 1-3.

3. Dhara, S.; Lu, C.-Y.; Magudapathy, P.; Huang, Y.-F.; Tu, W.-S.; Chen K.-H. Surface plasmon polariton assisted optical switching in noble bimetallic nanoparticle system. *Appl. Phys. Lett.* **2015**, *106*, 023101.

4. Dhara, S.; Lu, C.-Y.; Chen K.-H. Plasmonic switching in Au functionalized GaN nanowires in the realm of surface plasmon polatriton propagation :  A single nanowire switching device. *Plasmonics* **2014**, http://dx.doi.org/10.1007/s11468-014-9815-z

5. Atwater, H. A; Polman, A. Plasmonics for improved photovoltaic devices.  *Nature Mater.* **2010**, *9*, 205-213.

6. Lal, S., Clare, S. E.; Halas, N. J. Nanoshell-enabled photothermal cancer therapy: impending clinical impact. *Acc. Chem. Res.* **2008**, *41*, 1842–1851.

7. Juluri, B. K., Zheng, Y. B., Ahmed, D., Jensen, L.; Huang, T. J. Effects of geometry and composition on charge-induced plasmonic shifts in gold nanoparticles. *J. Phys. Chem. C* **2008**, *112*, 7309–7317.

8. Nie, S.; Emory, S. R. Probing single molecules and single nanoparticles by surface-enhanced Raman scattering. *Science* 1997, *275*, 1102-1106 .

9. Garg, P.; Dhara S. Single molecule detection using SERS study in PVP functionalized Ag nanoparticles. *AIP Conf. Proc.* **2013**, *1512*, 206-207.

10. Zhang, R.; Zhang, Y.; Dong, Z. C.; Jiang, S.; Zhang, C.; Chen, L. G.; Zhang, L.; Liao, Y.; Aizpurua, J.; Luo, Y.; Yang, J. L.; Hou, J. G. Chemical mapping of a single molecule by plasmon-enhanced Raman scattering. *Nature* **2013**, *498*, 82-86.

11. Kreibig, U.; Vollmer, M. Optical properties of metal clusters. *Springer series, Berlin*, **1995**.

12. Palpant, B.; Prével, B.; Lermé, J.; Cottancin, E.; Pellarin, M.; Treilleux, M.; Perez, A.; Vialle, J. L.; Broyer, M. Optical properties of gold clusters in the size range 2–4 nm. ***Phys. Rev. B*** **1998**, *57*, 1963-1970.

13. Lermé, J.; Palpant, B.; Prével, B.; Cottancin, E.; Pellarin, M.; Treilleux, M.; Vialle, J. L.; Perez, A.; Broyer, M. Optical properties of gold metal clusters: A time-dependent local-density-approximation investigation. *Eur. Phys. J. D* **1998**, *4*, 95-108.

14. Srivastava, S. K.; Gangopadhyay, P.; Amirthapandian, S.; Sairam, T. N.; Basu, J; Panigrahi, B.K.; Nair, K. G. M. Effects of high-energy Si ion-irradiations on optical responses of Ag metal nanoparticles in a SiO$_2$ matrix *Chem. Phys. Lett.* **2014** http://dx.doi.org/10.1016/ j.cplett.2014.05.059

15 Scholl, J. A.; Koh, A. L.; Dionne, J. A. Quantum plasmon resonances of individual metallic nanoparticles. *Nature* **2012**, *483*, 421-427.

16. Hosoya, Y.; Suga, T.; Yanagawa, T.; Kurokawa, Y. Linear and nonlinear optical properties of sol-gel-derived Au nanometer-particle-doped alumina. *J. Appl. Phys.* **1997**, *81*, 1475-1480.

17. Meldrum, A.; Boatner, L. A.; White C. W.; Ewing, R. C. Ion irradiation effects in nonmetals: formation of nanocrystals and novel microstructures. *Mater. Res. Innovat.* **2000**, *3*, 190-204.

18. Miotelo, A.; De Merchi, G.; Mattei, G.; Mazzoldi, P.; Sada, C. Clustering of gold atoms in ion-implanted silica after thermal annealing in different atmospheres. *Phy. Rev. B* **2001**, *63*, 075409.

19. Dai, Z.; Yamamoto, S.; Narumi, K.; Miyashita, A.; Naramoto, H. Gold nanoparticle fabrication in single crystal $SiO_2$ by MeV Au ion implantation and subsequent thermal annealing. *Nucl. Instr. Meth. Phys. Res. B* **1999**, *149*, 108-112.

20. Ila, D.; Williams, E. K.; Zimmerman, R. L.; Poker, D. B.; Hensley, D. K. Radiation induced nucleation of nanoparticles in silica. *Nucl. Instr. Meth. Phys. Res. B* **2000**, *166–167*, 845-850.

21. Dhara, S., Sundaravel, B.; Ravindran, T. R.; Nair, K. G. M.; David, C.; Panigrahi, B. K.; Magudapathy, P.; Chen K. H. 'Spillout' effect in gold nanoclusters embedded in c-$Al_2O_3$(0001) matrix. *Chem. Phys. Lett.* **2004**, *399*, 354-358.

22. Dhara, S.; Kesavamoorthy, R.; Magudapathy, P.; Premila, M.; Panigrahi, B. K.; Nair, K. G. M.; Wu, C.T.; Chen K. H. Quasiquenching size effects in gold nanoclusters embedded in silica matrix. *Chem. Phy. Lett .* **2003**, *370*, 254-260.

23. Raza, S.; Stenger, N.; Kadkhodazadeh, S.; Fischer, S. V.; Kostesha, N.; Jauho, A.-P.; Burrows, A,; Wubs, M.; Mortensen N. A. Blueshift of the surface plasmon resonance in silver nanoparticles studied with EELS. *Nanophotonics* **2013**, *2*, 131–138.

24. Palpant, B.; Portales, H.; Saviot, L.; Lerme, J.; Prevel, B.; Pellarin, M.; Duval, E.; Perez, A.; Broyer, M. Quadrupolar vibrational mode of silver clusters from plasmon-assisted Raman scattering. *Phys. Rev. B* **1999**, *60*, 17107-17111.

25. Saviot, L.; Murray, D. B.; de Lucas, M. Vibrations of free and embedded anisotropic elastic spheres: Application to low-frequency Raman scattering of silicon nanoparticles in silica. *Phys. Rev. B* **2004**, *69*, 113402-113404.

26. Ekardt, W. Work function of small metal particles: Self-consistent spherical jellium-background model. *Phys. Rev. B* **1984**, *29*, 1558-1564.

27. Wannier, G. H. Dynamics of band electrons in electric and magnetic fields. *Rev. Mod. Phys.* **1962**, *34*, 645-655.

28. Nenciu, G. Dynamics of band electrons in electric and magnetic fields: rigorous justification of the effective Hamiltonians. *Rev. Mod. Phys.* **1991**, *63*, 91-128.

29. Feldmann, J.; Leo, K.; Shah, J.; Miller, D. A. B.; Cunningham, J. E.; Meier, T.; von Plessen, G.; Schulze, A.; Thomas, P.; Schmitt-Rink, S. Optical investigation of Bloch oscillations in a semiconductor superlattice. *Phys. Rev. B* **1992**, *46*, 7252-7255.

30. Hawton, M.; Dignam, M. M. Infinite-order excitonic Bloch equations for asymmetric nanostructures. *Phys. Rev. Lett.* **2003**, *91*, 267402.

31. Bruggeman, D. A. G. Calculation of various physical constants of heterogeneous substances. *Ann. Phys.* (*Leipzig*) **1935**, *24*, 636-679.

32. Tiggesbäumker, J.; Köller, L.; Meiwes-Broer, K. H.; Liebsch, A. Blue shift of the Mie plasma frequency in Ag clusters and particles. *Phys. Rev. A* 1993, *48*, R1749 –R1752.

33. Fedrigo, S.; Harbich, W.; Buttet, J. Collective dipole oscillations in small silver clusters embedded in rare-gas matrices. *Phys. Rev. B* **1993**, *47*, 10 706.

34. Charl´e, K.P., Schulze W., Winter, B. The size dependent shift of the surface plasmon absorption band of small spherical metal particles. *Z. Phys. D* **1989**, *12*, 471–475.

35. Genzel, L.; Martin, T. P.; Kreibig, U. Dielectric function and plasma resonances of small metal particles. *Z. Phys. B* **1975**, *21*, 339-346.

36. Cottancin, E.; Lerme´, J.; Gaudry, M.; Pellarin, M.; Vialle, J.-L.; Broyer, M.; Pre´vel, B.; Treilleux, M.; Me´linon, P. Size effects in the optical properties of $Au_nAg_n$ embedded clusters *Phys. Rev B* **2000**, *62*, 5179-5185.

37. Thongrattanasiri, S., Manjavacas, A.; García de Abajo, F. J. Quantum finite-size effects in graphene plasmons. *ACS Nano* **2012**, *6*, 1766–1775.

38. Bailo, E.; Deckert, V. Tip-enhanced Raman scattering. *Chem. Soc. Rev.* **2008**, *37*, 921–930.

39. Gaudry, M.; Lerme´, J.; Cottancin, E.; Pellarin, M.; Vialle, J. -L.; Broyer, M.; Pre´vel, B.; Treilleux, M.; Me´linon, P. Optical properties of $(Au_xAg_{1-x})_n$ clusters embedded in alumina: Evolution with size and stoichiometry. *Phys. Rev. B* **2001**, *64*, 085407.

40. Ripken, K. Optical-Constants of Au, Ag and their alloys in energy region from 2.4 to 4.4 eV *Z. Phys.* **1972**, *250*, 228-234.

41. Nilsson, P. O. Electronic structure of disordered alloys: Optical and photoemission measurements on Ag-Au and Cu-Au alloys. *Phys. Kondens. Mater.* **1970,** *11*, 1-18.

42. Rakić, A. D.; Djurišić, A. B.; Elazar, J. M.; Majewski M. L. Optical properties of metallic films for vertical-cavity optoelectronic devices. *Appl. Opt.* **1998**, *37*, 5271-5283.

**Figure Captions :**

**Fig. 1.** SPR frequencies for embedded Au nanoclusters in c-Al$_2$O$_3$ annealed at 1273 and 1473 K showing redshift with decreasing fluence. Dashed vertical line is a guide to eye for the observed red shift of the SPR frequency with decreasing fluence (Ref [21]; *Applied for Copyright @ Elsevier*).

**Fig. 2.** a) Size dependence of the SPR frequency for the samples annealed at 1273K and 1473K (unfilled symbols) and estimated values from TDLDA calculations (filled symbols) (Ref [13]) corresponding to *r*=0 and 1 a.u. are presented. Experimental values show a red shift with decreasing cluster size (Ref [21]; *Applied for Copyright @ Elsevier*). ). b) Schematic of two region dielectric model with nanoclusters radius *R* and screening length *r*. The reduction in the electron density as a function of *r*, *n(r)* is shown to reduce with decreasing *R* (spillout effect).

**Fig. 3.** a) –d) Photoabsorption spectra of free gold clusters within the two-region dielectric model, for different thicknesses of the skin region of reduced polarizability. Thick line curves: *d* =2 a.u.; dashed line curves: *d* = 1 a.u.; thin line curves: *d* = 0. The short vertical line in d) indicates the dipolar Mie resonance energy in the large-particle limit (Ref [13]; *Applied for Copyright @ EDP Sciences, Springer-Verlag*).

**Fig. 4.** The size evolution of the surface plasmon energy of gold clusters within different models. Black triangles: TDLDA results obtained with an outer vacuum "rind" at the metal/alumina matrix interface. $d_m$ is the thickness of the outer rind. Empty triangles: TDLDA results for free (upper curve, $d_m = \infty$) and alumina matrix-embedded clusters (lower curve, $d_m = 0$). Crosses: TDLDA results with a homogeneous surrounding matrix characterized by a constant dielectric function $\varepsilon_m =$ 2. In all calculations the thickness *d* of the inner skin of reduced polarizability is equal to 2 a.u. Black squares: experimental results (Ref [12]). The short horizontal line at 2.25 eV indicates the surface plasmon energy in the large-particle limit for fully-embedded AuN clusters. The circles

correspond to experimental (black) and TDLDA (empty) results for alumina matrix-embedded $Ag_N$ clusters (model parameters: $d = 2$ a.u. and $d_m = 4$ a.u.) (Ref [13]; *Applied for Copyright @ EDP Sciences, Springer-Verlag*).

**Fig. 5.** Size evolution of the maximum of the Mie-resonance peak for free $Ag_N$ (triangles) and $Ag_N^+$ (circles) clusters within the two-region dielectric model, for different values of the thickness parameter $d$. Black squares: experimental data on $Ag_N^+$ clusters (Ref [32]). The short horizontal line at 3.41 eV is the Mie frequency in the large-particle limit. (Ref [1]; *Applied for Copyright @ American Physical Society*).

**Fig. 6.** a) Surface plasmon resonance peaks for the annealed (typically for 1173 K) samples grown at various irradiation fluences showing the blue shift with decreasing fluence. The spectra are vertically shifted for clarity. b) Spectral dependence of the real and imaginary components of the complex dielectric function corresponding to the core electrons for gold metal. Im[$\varepsilon_d(\omega)$] is plotted with $\varepsilon_d(\omega) = \exp(\text{const.}(\omega-2.4))$ (near L point in the Birllouin zone: 1.7 eV$\leq\omega\leq$2.4 eV) and $\varepsilon_d(\omega) = (\omega-2.4)/\omega^2$ (near point in the Birllouin zone : 2.4 eV $\leq\omega\leq$3.3 eV) (Ref [11]). Re[$\varepsilon_d(\omega)$] is calculated using Kramers–Kronig relationship (Ref [11]).for the corresponding ranges. The vertical line corresponds to the Mie frequency for large free clusters. (Ref [22]; *Applied for Copyright @ Elsevier*).

**Fig. 7.** The size evolution of the peak plasmon maximum within different models for particles embedded in an alumina matrix. a) Classical results for $(Au_{0.5}Ag_{0.5})_n$ clusters (dots); (a) and (b) TDLDA results (solid line) and experimental results (black triangles) for $(Au_{0.5}Ag_{0.5})_n$ clusters; (b) TDLDA results (dashed line) and experimental results (open triangles) for $Ag_n$ clusters; TDLDA results (dotted line) and experimental results (open rhomb) for $Au_n$ clusters. (Ref [36]; *Applied for Copyright @ American Physical Society*).

**Fig. 8.** Correlating Ag nanoparticle geometry with plasmonic EELS data. a) Collection of normalized, deconvoluted EELS data from particles ranging from 11 nm to 1.7 nm in diameter and the corresponding STEM image of each specimen. The electron beam was directed onto the edge of the particles so that only the surface resonance is shown. b) Plot of the surface plasmon resonance energy versus particle diameter, with the inset depicting bulk resonance energies. Horizontal error bars indicating 95% confidence intervals were generated with a curve fitting and bootstrapping technique (see Methods). Vertical error bars are contained within the size of the data points (Ref [15]; *Applied for Copyright @ Nature Publications*).

**Fig. 9.** Comparison of experimental data with quantum theory. Experimental, EELS-determined localized surface plasmon resonance energies of various Ag particle diameters are overlaid on the absorption spectra generated from the analytic quantum permittivity model a) and the DFT-derived permittivity model b). The experimental bulk resonance energies are also included (grey dots) along with the theory prediction (grey line). Classical Mie theory peak prediction is given by the dashed white line. The experimental data begin to deviate significantly from classical predictions for particle diameters smaller than 10nm. Horizontal error bars represent 95% confidence intervals, as calculated through curve fitting and bootstrapping techniques (Ref [15]; *Applied for Copyright @ Nature Publications*).

**Fig. 10.** Aberration-corrected STEM images of Ag nanoparticles with diameters (a) 15.5 nm, (b) 10 nm, and (c) 5.5 nm, and normalized raw EELS spectra of similar-sized Ag nanoparticles (d-f). The EELS measurements are acquired by directing the electron beam to the surface of the particle (Ref. [23]; *Applied for Copyright @ Science Wise Publishing and De Gruyter*).

**Fig. 11.** Nanoparticle SPR energy as a function of the particle diameter. The dots are EELS measurements taken at the surface of the particle and analyzed using the Reflection Tail (RT)

method, and the lines are theoretical predictions. We use parameters from Ref [42]: $\hbar\omega_p$=8.282 eV, $\hbar\gamma$=0.048 eV, $n_0 = 5.9 \times 10^{28}$ m$^{-3}$ and $v_F = 1.39 \times 10^6$ m/s. From the average large-particle ($2R > 20$ nm) resonances we determine $\varepsilon_B = 1.53$ (Ref. [23]; *Applied for Copyright @ Science Wise Publishing and De Gruyter*).

**Fig. 12.** (Top panel) Optical absorption spectra of Ag ion-implanted ($5 \times 10^{16}$ ions cm$^{-2}$) SiO$_2$ samples followed by Si ion irradiations (ion dose as indicated) are displayed. Blue shifts of SPR peak and systematic decrease of the resonance intensity with increase of Si ion fluence may be observed. (Bottom panel) Graphs A & B display the optical absorption spectra of Ag particles in SiO$_2$ samples before & after Si ion-irradiations ($5 \times 10^{15}$ ions cm$^{-2}$), respectively. Ag ion-implantation fluence is $1 \times 10^{16}$ ions cm$^{-2}$. (Inset) Graph C displays the optical absorption spectra of Ag particles in SiO$_2$ samples after Si ion-irradiations ($5 \times 10^{15}$ ions cm$^{-2}$). Ag ion implantation fluence is 3x1016 ions cm-2. Subsequent annealing (400 $^o$C, 1 hr) the SiO$_2$ sample in an inert gas atmosphere brings back the SPR characteristics of larger Ag nanoparticles (see graph D). Quantum nature of surface plasmon resonances in fine Ag particles vanishes, revealing ripening of the Ag particles on thermal annealing (Ref [14]; *Applied for Copyright @ Elsevier*).

**Fig. 13.** Spectral dependence of the imaginary part of the dielectric function of the core electrons in the (Au$_x$Ag$_{1-x}$) alloy, with $x = 0, 0.25, 0.5, 0.75, 1$ from top to bottom, within two different hypotheses (see text), a) model 1 b) and model 2 (Ref [39]; *Applied for Copyright @ American Physical Society*).

**Fig. 14.** Size evolution of the peak plasmon maximum for (Au$_x$Ag1-x)$_n$ clusters ($x = 0$ to 1) embedded in an alumina matrix, assuming two hypotheses for the dielectric function associated with the core electrons (see text): model 1 a) and model 2 b). TDLDA results, open symbols connected with lines; experimental results, large black symbols. Owing to the rather tiny effects (all the values lie in the energy range 2.3–3 eV) a finer increment $\Delta\omega$, ten times smaller than the increment involved in the spectra in Fig. 6 of Ref [39], has been used over the resonance spectral range in order to unambiguously determine the maximum locations, and thus the size evolution (Ref [39]; *Applied for Copyright @ American Physical Society*).

**Figures :**

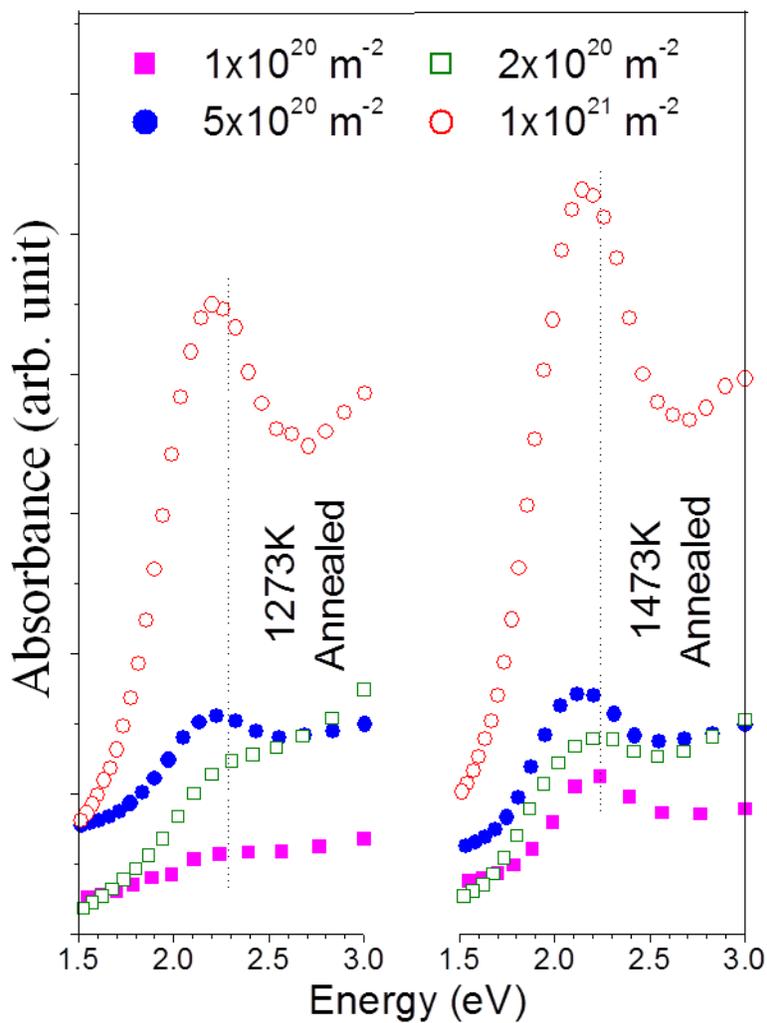

**Fig. 1.** SPR frequencies for embedded Au nanoclusters in c-$Al_2O_3$ annealed at 1273 and 1473 K showing redshift with decreasing fluence. Dashed vertical line is a guide to eye for the observed red shift of the SPR frequency with decreasing fluence (Ref [21]; *Applied for Copyright @ Elsevier*).

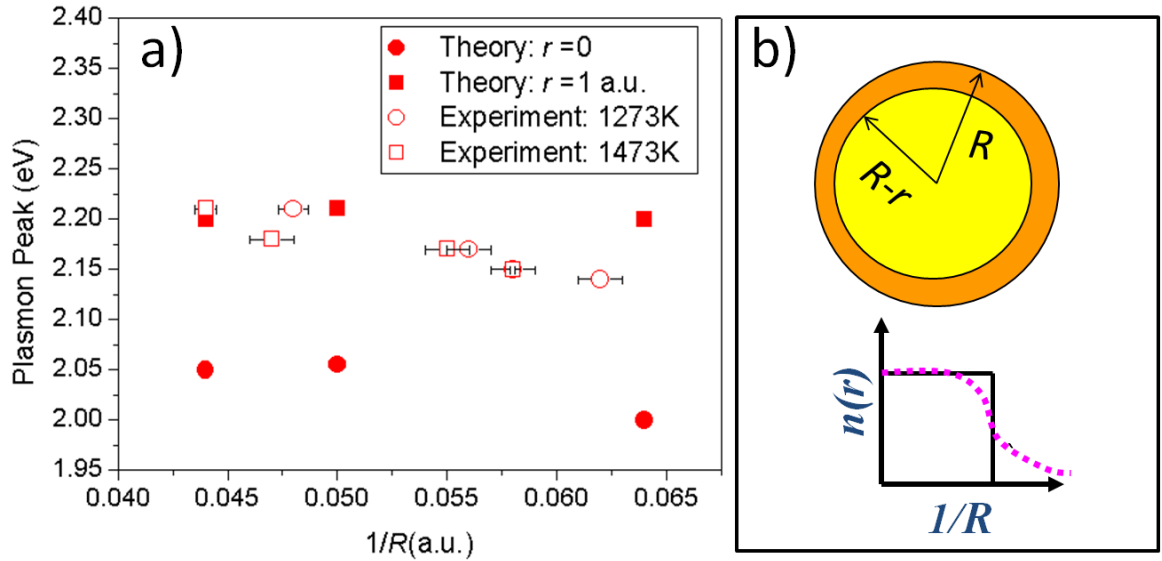

**Fig. 2.** a) Size dependence of the SPR frequency for the samples annealed at 1273K and 1473K (unfilled symbols) and estimated values from TDLDA calculations (filled symbols) (Ref [13]) corresponding to *r*=0 and 1 a.u. are presented. Experimental values show a red shift with decreasing cluster size (Ref [21]; *Applied for Copyright @ Elsevier*). b) Schematic of two region dielectric model with nanoclusters radius *R* and screening length *r*. The reduction in the electron density as a function of *r*, *n(r)* is shown to reduce with decreasing *R* (spillout effect).

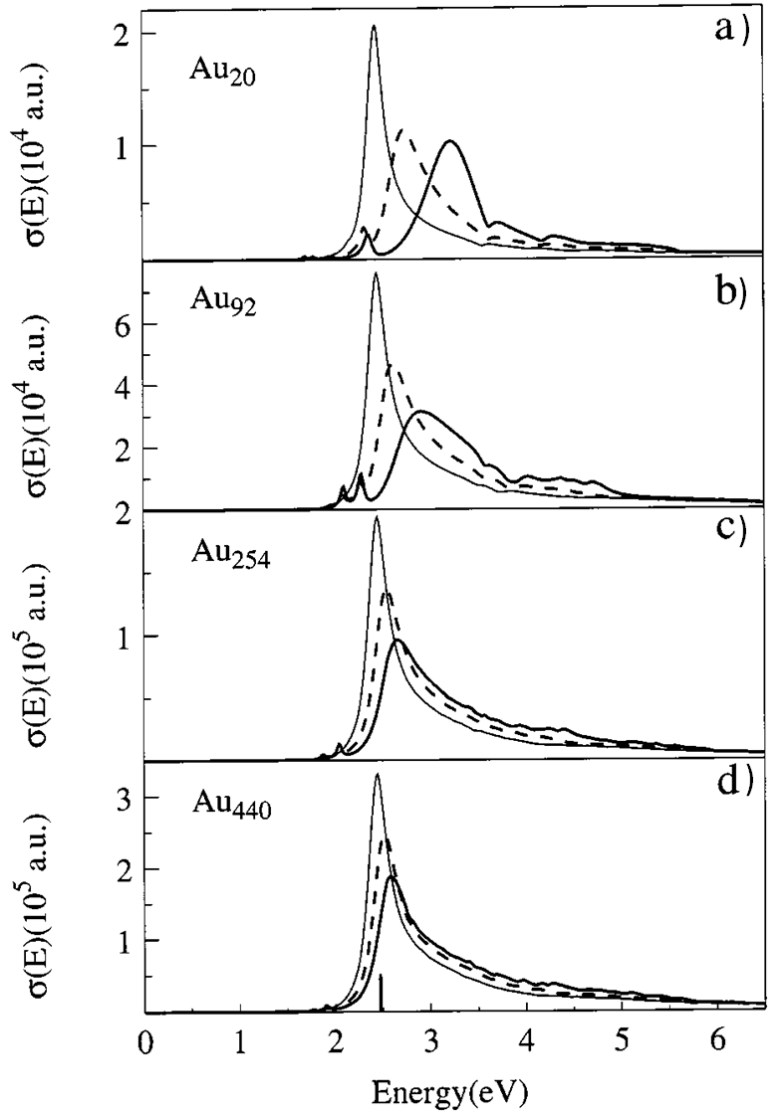

**Fig. 3.** a) –d) Photo-absorption spectra of free gold clusters within the two-region dielectric model, for different thicknesses of the skin region of reduced polarizability. Thick line curves: $d = 2$ a.u.; dashed line curves: $d = 1$ a.u.; thin line curves: $d = 0$. The short vertical line in d) indicates the dipolar Mie resonance energy in the large-particle limit (Ref [13]; *Applied for Copyright @ EDP Sciences, Springer-Verlag*).

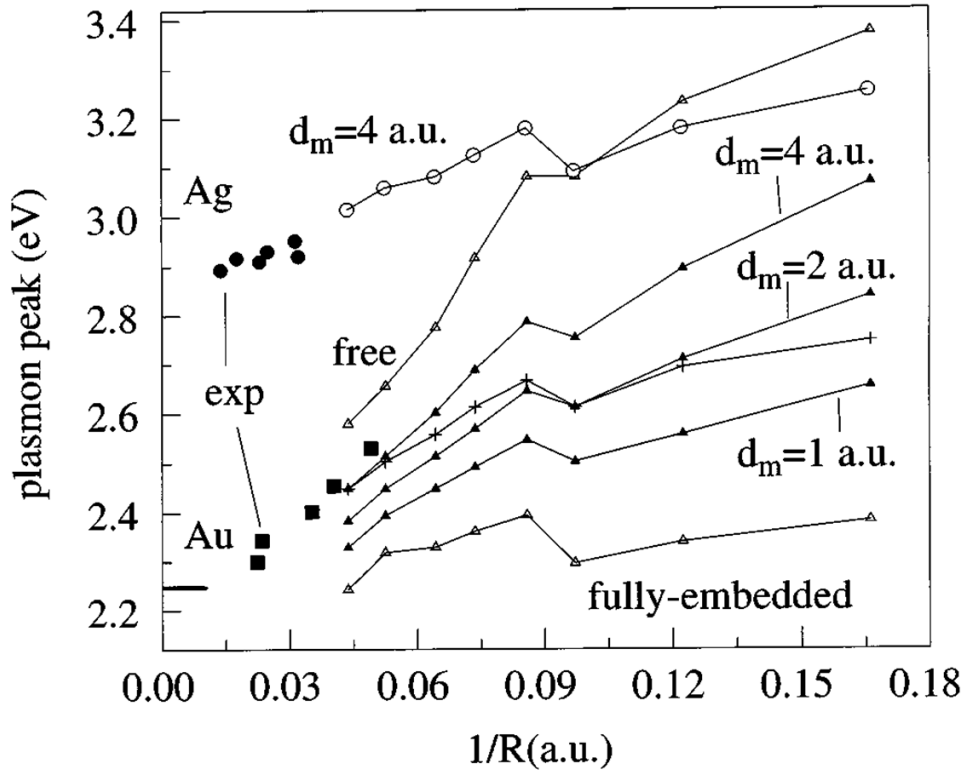

**Fig. 4.** The size evolution of the surface plasmon energy of gold clusters within different models. Black triangles: TDLDA results obtained with an outer vacuum "rind" at the metal/alumina matrix interface. dm is the thickness of the outer rind. Empty triangles: TDLDA results for free (upper curve, $d_m = \infty$) and alumina matrix-embedded clusters (lower curve, $d_m = 0$). Crosses: TDLDA results with a homogeneous surrounding matrix characterized by a constant dielectric function $\varepsilon_m = 2$. In all calculations the thickness $d$ of the inner skin of reduced polarizability is equal to 2 a.u. Black squares: experimental results (Ref [12]). The short horizontal line at 2.25 eV indicates the surface plasmon energy in the large-particle limit for fully-embedded AuN clusters. The circles correspond to experimental (black) and TDLDA (empty) results for alumina matrix-embedded AgN clusters (model parameters: $d = 2$ a.u. and $d_m = 4$ a.u.) (Ref [13]; *Applied for Copyright @ EDP Sciences, Springer-Verlag*).

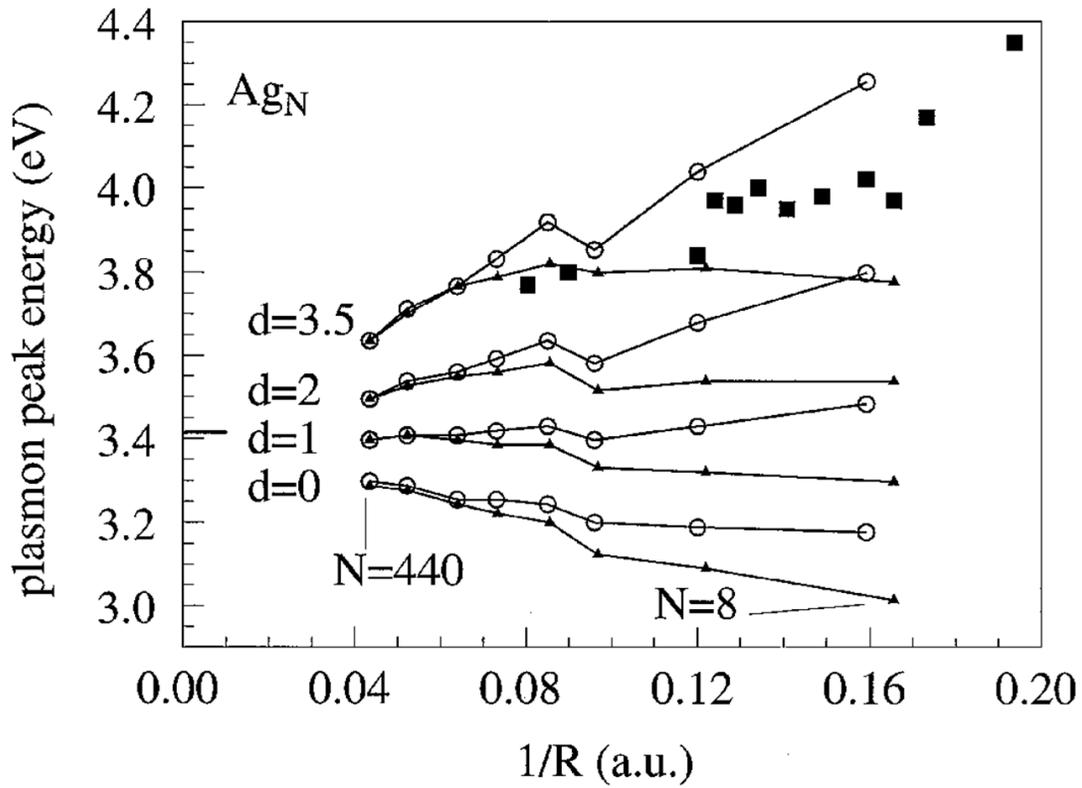

**Fig. 5.** Size evolution of the maximum of the Mie-resonance peak for free $Ag_N$ (triangles) and $Ag_N^+$ (circles) clusters within the two-region dielectric model, for different values of the thickness parameter *d*. Black squares: experimental data on $Ag_N^+$ clusters (Ref. [32]). The short horizontal line at 3.41 eV is the Mie frequency in the large-particle limit. (Ref [1]; *Applied for Copyright @ American Physical Society*).

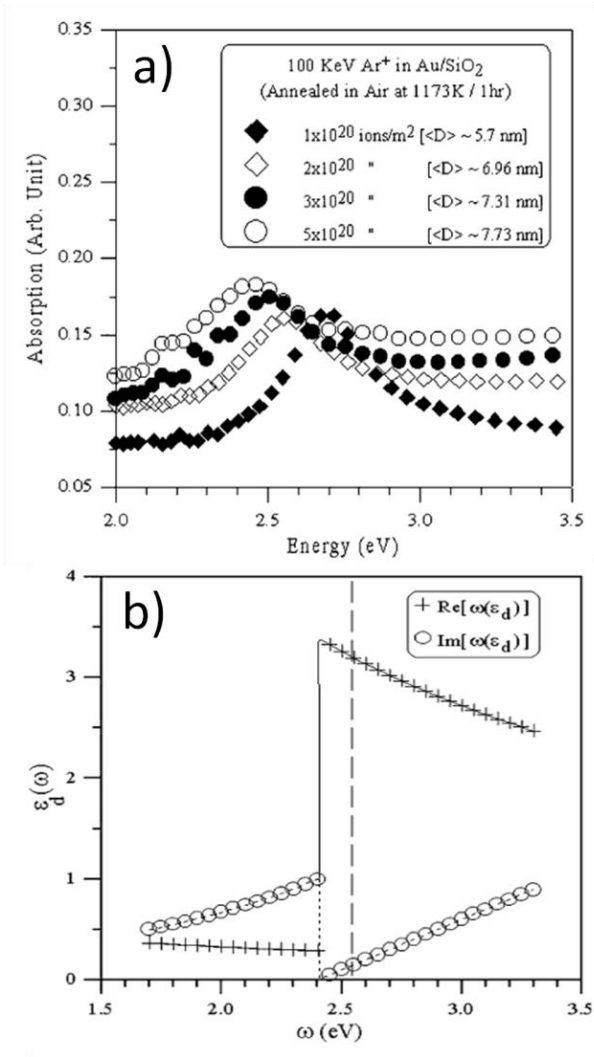

**Fig. 6.** a) Surface plasmon resonance peaks for the annealed (typically for 1173 K) samples grown at various irradiation fluences showing the blue shift with decreasing fluence. The spectra are vertically shifted for clarity. b) Spectral dependence of the real and imaginary components of the complex dielectric function corresponding to the core electrons for gold metal. Im[$\varepsilon_d(\omega)$] is plotted with $\varepsilon_d(\omega)$ = exp(const.($\omega$-2.4)) (near L point in the Birllouin zone: 1.7 eV≤$\omega$≤2.4 eV) and $\varepsilon_d(\omega)$ = ($\omega$-2.4)/$\omega^2$ (near point in the Birllouin zone : 2.4 eV ≤$\omega$≤3.3 eV) (Ref [11]). Re[$\varepsilon_d(\omega)$] is calculated using Kramers–Kronig relationship (Ref [11]).for the corresponding ranges. The vertical line corresponds to the Mie frequency for large free clusters. (Ref [22]; *Applied for Copyright @ Elsevier*).

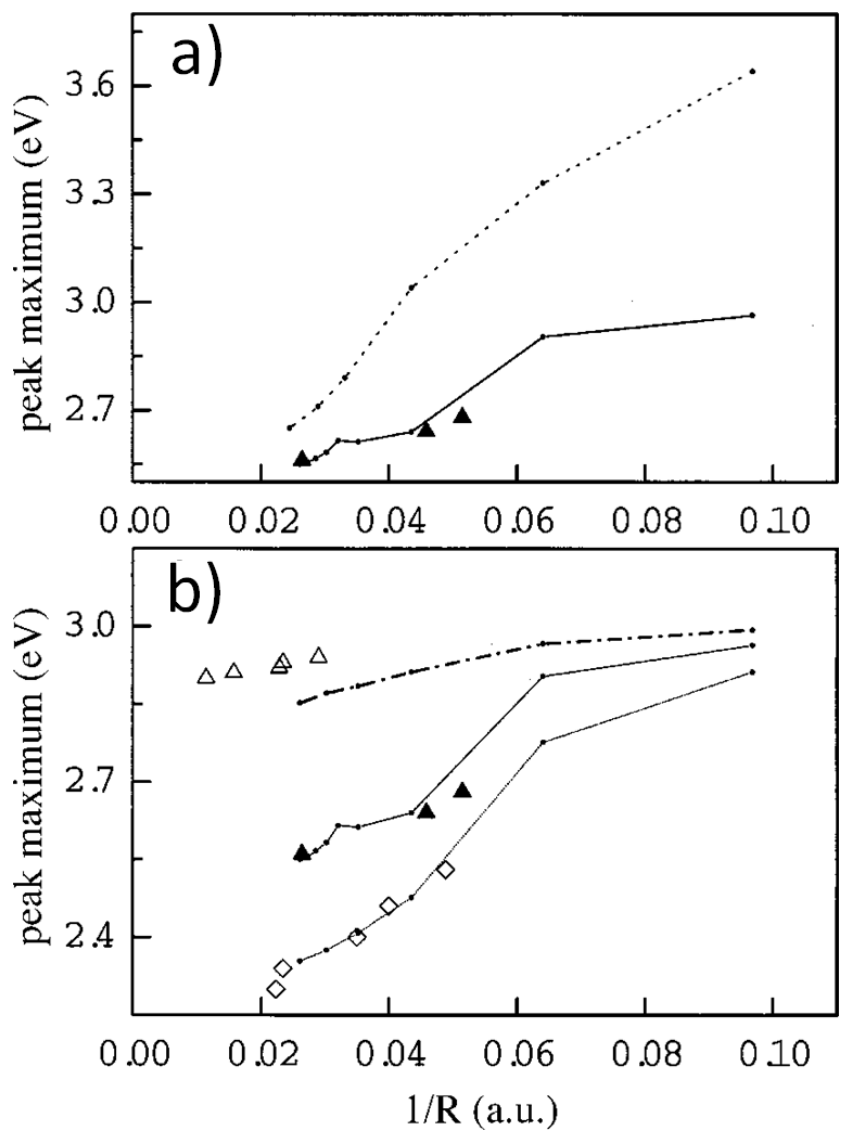

**Fig. 7.** The size evolution of the peak plasmon maximum within different models for particles embedded in an alumina matrix. a) Classical results for $(Au_{0.5}Ag_{0.5})_n$ clusters (dots); (a) and (b) TDLDA results (solid line) and experimental results (black triangles) for $(Au_{0.5}Ag_{0.5})_n$ clusters; (b) TDLDA results (dashed line) and experimental results (open triangles) for $Ag_n$ clusters; TDLDA results (dotted line) and experimental results (open rhomb) for $Au_n$ clusters. (Ref [36]; *Applied for Copyright @ American Physical Society*).

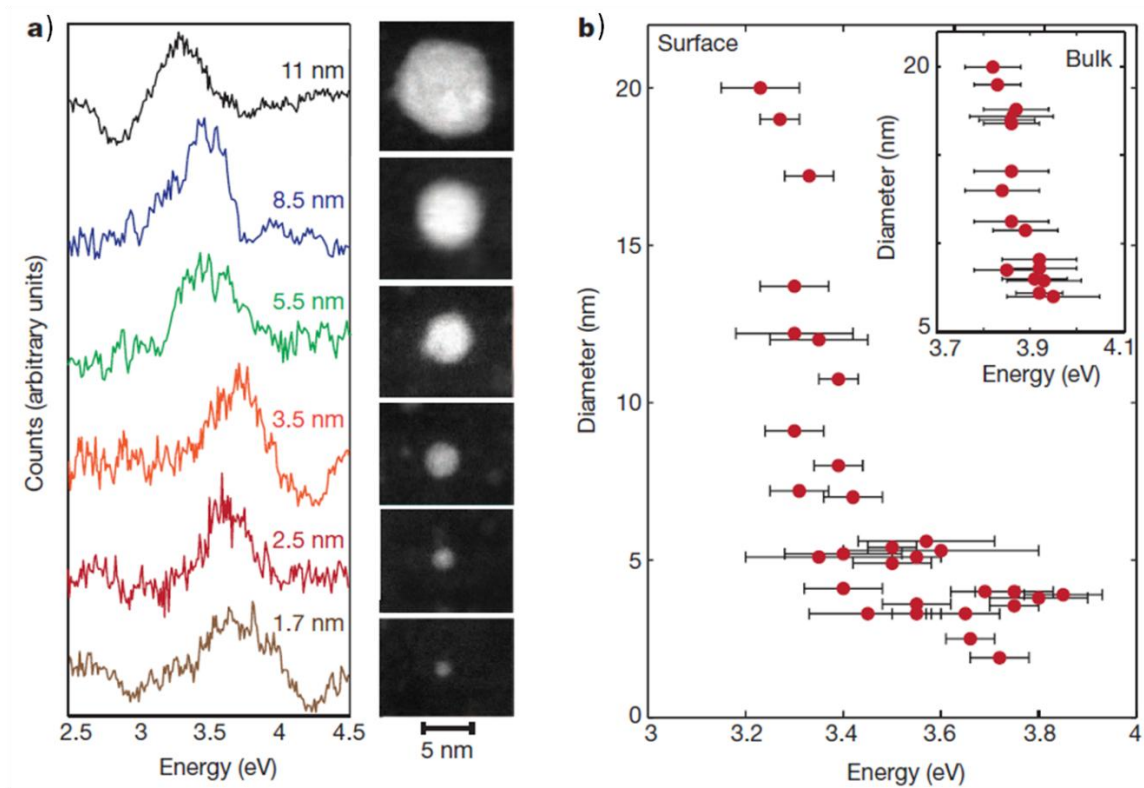

**Fig. 8.** Correlating Ag nanoparticle geometry with plasmonic EELS data. a) Collection of normalized, deconvoluted EELS data from particles ranging from 11 nm to 1.7 nm in diameter and the corresponding STEM image of each specimen. The electron beam was directed onto the edge of the particles so that only the surface resonance is shown. b) Plot of the surface plasmon resonance energy versus particle diameter, with the inset depicting bulk resonance energies. Horizontal error bars indicating 95% confidence intervals were generated with a curve fitting and bootstrapping technique (see Methods). Vertical error bars are contained within the size of the data points (Ref [15]; *Applied for Copyright @ Nature Publications*).

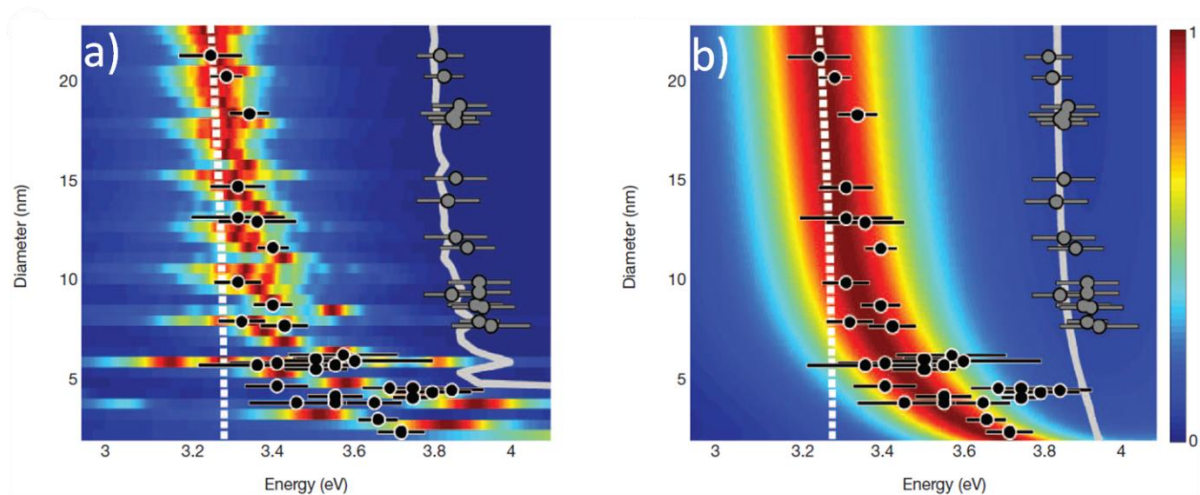

**Fig. 9.** Comparison of experimental data with quantum theory. Experimental, EELS-determined localized surface plasmon resonance energies of various Ag particle diameters are overlaid on the absorption spectra generated from the analytic quantum permittivity model a) and the DFT-derived permittivity model b). The experimental bulk resonance energies are also included (grey dots) along with the theory prediction (grey line). Classical Mie theory peak prediction is given by the dashed white line. The experimental data begin to deviate significantly from classical predictions for particle diameters smaller than 10nm. Horizontal error bars represent 95% confidence intervals, as calculated through curve fitting and bootstrapping techniques (Ref [15]; *Applied for Copyright @ Nature Publications*).

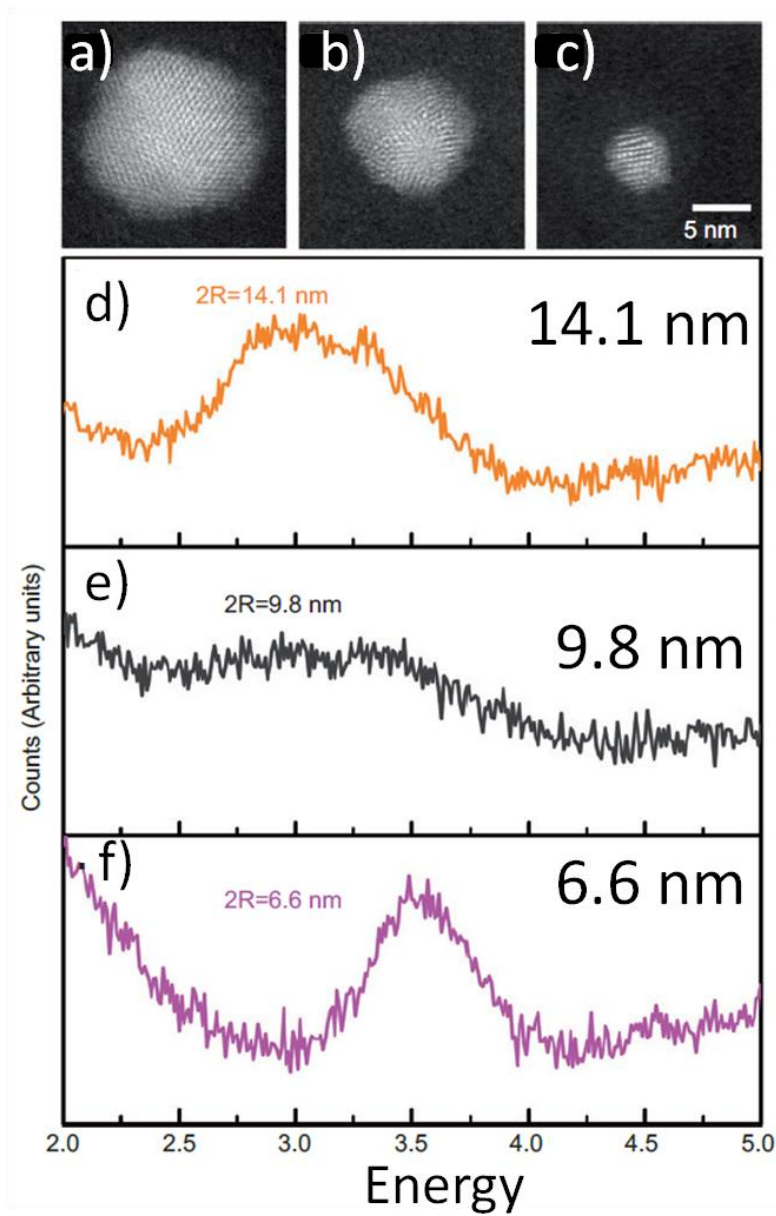

**Fig. 10.** Aberration-corrected STEM images of Ag nanoparticles with diameters (a) 15.5 nm, (b) 10 nm, and (c) 5.5 nm, and normalized raw EELS spectra of similar-sized Ag nanoparticles (d-f). The EELS measurements are acquired by directing the electron beam to the surface of the particle (Ref. [23]; *Applied for Copyright @ Science Wise Publishing and De Gruyter*).

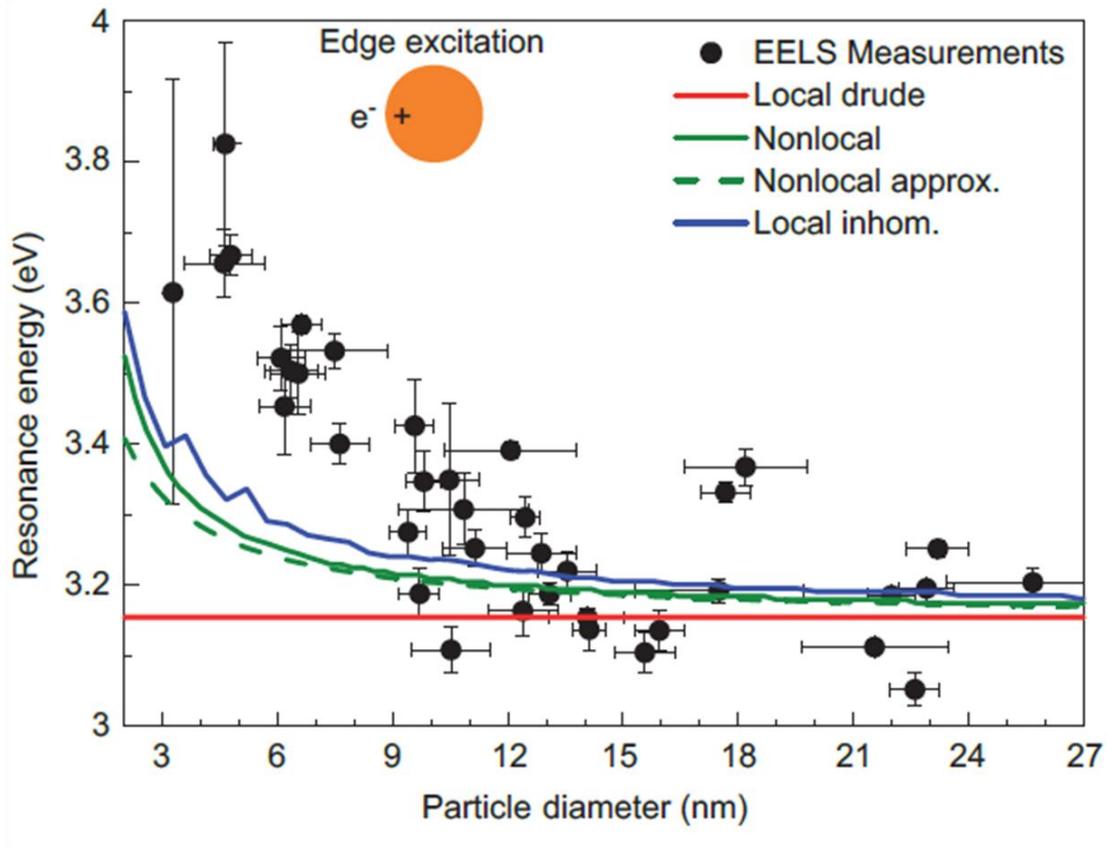

**Fig. 11.** Nanoparticle SPR energy as a function of the particle diameter. The dots are EELS measurements taken at the surface of the particle and analyzed using the Reflection Tail (RT) method, and the lines are theoretical predictions. We use parameters from Ref [42]: $\hbar\omega_p$=8.282 eV, $\hbar\gamma$=0.048 eV, $n_0 = 5.9 \times 10^{28}$ m$^{-3}$ and $v_F = 1.39 \times 10^6$ m/s. From the average large-particle ($2R > 20$ nm) resonances we determine $\varepsilon_B = 1.53$ (Ref. [23]; *Applied for Copyright @ Science Wise Publishing and De Gruyter*).

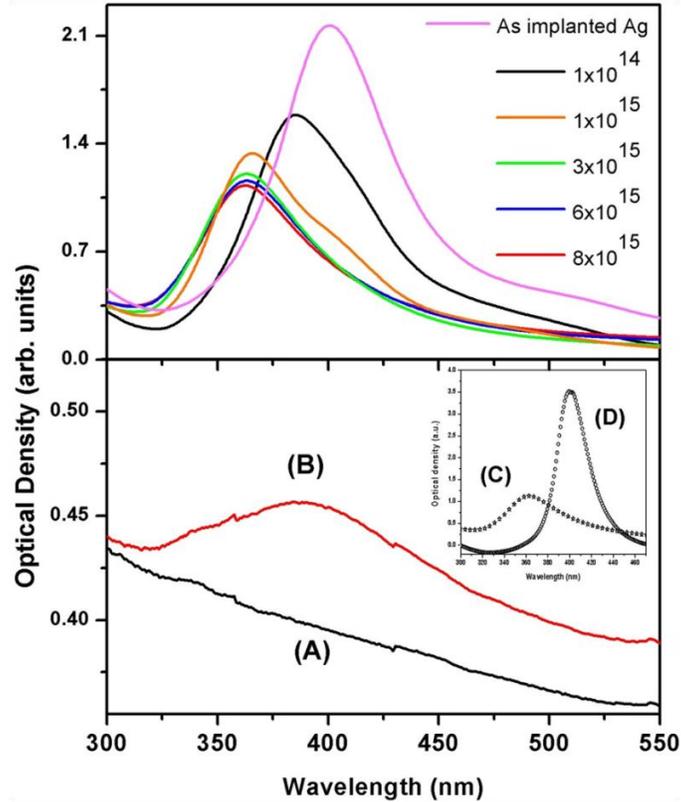

**Fig. 12.** (Top panel) Optical absorption spectra of Ag ion-implanted ($5 \times 10^{16}$ ions cm$^{-2}$) SiO$_2$ samples followed by Si ion irradiations (ion dose as indicated) are displayed. Blue shifts of SPR peak and systematic decrease of the resonance intensity with increase of Si ion fluence may be observed. (Bottom panel) Graphs A & B display the optical absorption spectra of Ag particles in SiO$_2$ samples before & after Si ion-irradiations ($5 \times 10^{15}$ ions cm$^{-2}$), respectively. Ag ion-implantation fluence is $1 \times 10^{16}$ ions cm$^{-2}$. (Inset) Graph C displays the optical absorption spectra of Ag particles in SiO$_2$ samples after Si ion-irradiations ($5 \times 10^{15}$ ions cm$^{-2}$). Ag ion implantation fluence is 3x1016 ions cm-2. Subsequent annealing (400 °C, 1 hr) the SiO$_2$ sample in an inert gas atmosphere brings back the SPR characteristics of larger Ag nanoparticles (see graph D). Quantum nature of surface plasmon resonances in fine Ag particles vanishes, revealing ripening of the Ag particles on thermal annealing (Ref [14]; *Applied for Copyright @ Elsevier*).

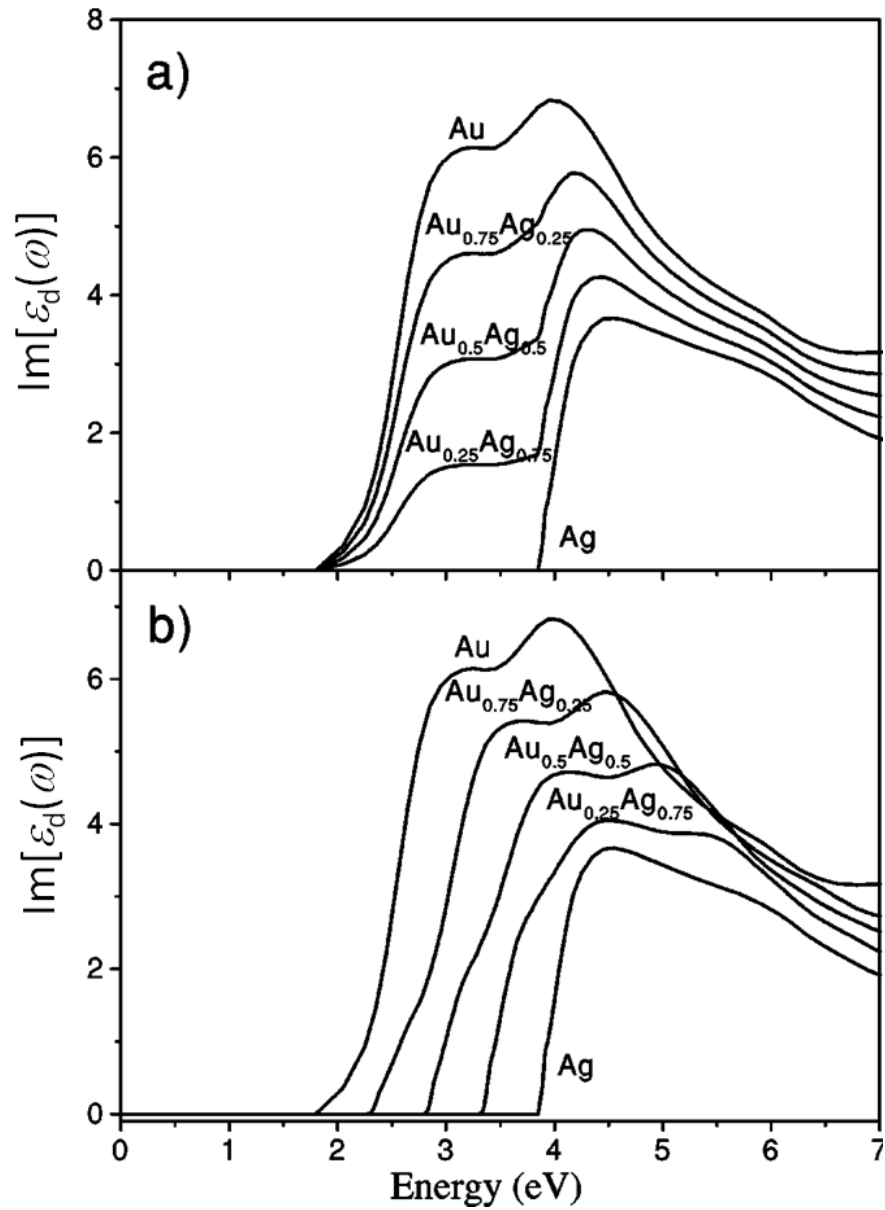

**Fig. 13.** Spectral dependence of the imaginary part of the dielectric function of the core electrons in the ($Au_xAg_{1-x}$) alloy, with $x = 0, 0.25, 0.5, 0.75, 1$ from top to bottom, within two different hypotheses (see text), a) model 1 b) and model 2 (Ref [39]; *Applied for Copyright @ American Physical Society*).

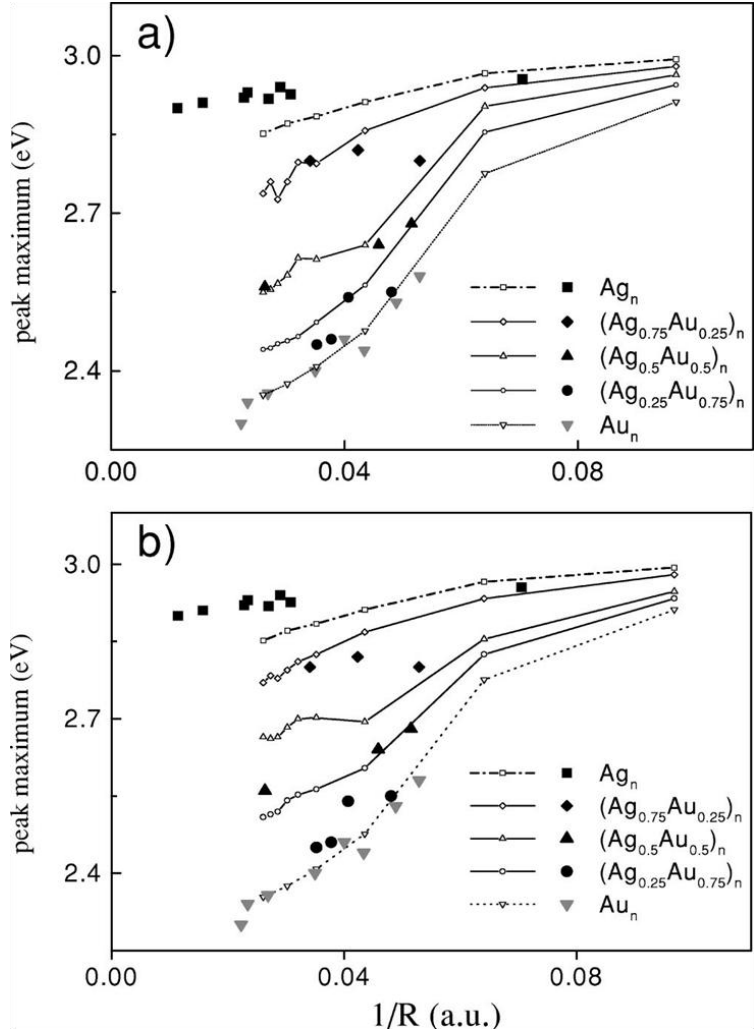

**Fig. 14.** Size evolution of the peak plasmon maximum for $(Au_xAg_{1-x})_n$ clusters ($x$ = 0 to 1) embedded in an alumina matrix, assuming two hypotheses for the dielectric function associated with the core electrons (see text): model 1 a) and model 2 b). TDLDA results, open symbols connected with lines; experimental results, large black symbols. Owing to the rather tiny effects (all the values lie in the energy range 2.3–3 eV) a finer increment $\Delta\omega$, ten times smaller than the increment involved in the spectra in Fig. 6 of Ref [39], has been used over the resonance spectral range in order to unambiguously determine the maximum locations, and thus the size evolution (Ref [39]; *Applied for Copyright @ American Physical Society*).